# IoT Smart City Architectures: an Analytical Evaluation

Mahdi Fahmideh
University of Technology Sydney (UTS)
Sydney, Australia
mahdi.fahmideh@uts.edu.au

Didar Zowghi
University of Technology Sydney (UTS)
Sydney, Australia
didar.zowghi@uts.edu.au

*Abstract*—while several IoT architectures have been proposed for enabling smart city visions, not much work has been done to assess and compare these architectures. By applying our proposed evaluation framework that incorporates a variety of 33 criteria, this paper presents a comparative analysis of nine existing well-known IoT architectures. The results of the analysis highlight the strengths and weaknesses of these architectures and give insight to city leaders, architects, and developers aiming at selecting the most appropriate architecture or their combination that may fit their own specific smart city development scenario.

## I. INTRODUCTION

The IoT initiatives are considered as a key technical enabler of smart city vision to enhance the quality of citizen's life [1]. Many valuable IoT architectures have already been proposed to operationalize smart city vision. This has raised the need to appraise and compare existing architectures through which one can identify their characteristics, strengths, weaknesses, and determine the one that seems to be most suitable to adopt in a particular smart city development scenario. There is a clear lack of an evaluation framework to help smart city developers to perform an effective and reliable assessment of existing IoT smart city architectures. We present the results of our analytical review of a number of the existing IoT architectures using our proposed evaluation framework. Our research has been carried out using the following steps: (i) selecting a sample IoT architectures, (ii) compiling a set of criteria derived from the IoT, smart city and software engineering literature, and (iii) appraising the quality of the selected architectures in the view of the criteria, and (iv) documenting the analytical results and observations from the evaluation. The evaluation results give insights to researcher and developers on the gaps that are yet to be addressed by existing architectures, constructing new research opportunities. Thus, this research makes contributions to the smart city architecture and IoT literature in a two ways: (i) the proposed evaluation framework serving as a valuable tool to examine, compare, prioritize, and select architectures attuned with smart city development scenario's goals, and (ii) using the presented evaluation results as an entry point for the purpose of situational smart city architecture design thus useful capabilities from the existing architectures are selected and combined to create bespoke architectures that fit the requirements of a smart city scenario.

This paper is organized as follows: Section I presents an introduction of a number of the existing IoT architectures. Section II presents the proposed evaluation framework to assess these architectures, followed by Section III that reports the evaluation results of the architectures. This continues with a brief discussion on the reliability of our evaluation results. After reviewing the related work in Section V, the paper concludes in Section VI.

## II. REVIEW OF SMART CITY ARCHITECTURES

This section presents an overview of the selected IoT architectures highlighting their key features. The key reasons that influenced the selection of these architectures were: (i) the availability of proper documentation to conduct a thorough review, (ii) a clear description of goals, architecture layers, and suggested, if any, guidelines or activities to address unique requirements of smart city development and maintenance. Among plethora of proposed architectures in academia and industry, we identified nine examples. These are BSI, TSB, QGC, FIWARE, IBM, ISO/IEC 30182, SOA, Cisco, and ESPRESSO, which are briefly reviewed in the following sub sections.

### A. BSI (British standard institute)

BSI, a standard architecture, offered by the UK Government to help city leaders respond to challenges and to support the development of smarter cities [2]. It offers advice identified from a wide range of public, private and voluntary practitioners engaged in facilitating the UK smart city services. The architecture emphasizes on the role of governance, culture, business model innovation, and stakeholders in the development, delivery, and use of city services. Finally, BSI provides a generic development methodology including five phases namely: plan, initiate, deliver, consolidate, and transform to build smart services.

### B. TSB (Technology Strategy Board)

The TSB, also called Innovate UK, is another UK government initiative for the development of smart cities [3]. This architecture aims to integrate technologies and service applications in public safety, transport, health, and sustainable energy to make a smarter city. A key concern in the TSB is cities' visions on environmental sustainability in terms of emissions reduction targets for the local area. The TSB's defines layers namely: systems application layer, platform layer, infrastructure layer, and organization layer.

*C. OGC (Open Geospatial Consortium)*

The OGC is an international consortium of companies, government agencies, and universities cooperating in the development of publicly available geospatial standards [4]. The platform provides a spatial information framework for urban spatial intelligence based on open standards such as CityGML [5], IndoorGML [6], and Augmented Reality Markup Language 2.0 [7]. The standard architecture provides a basis for integrating geographical information systems features, sensor observations, and social media in the support of city governance and services. Furthermore, the OGC provides open standards for mobile location communication, 3D urban models, building information models, augmented reality, and sensor webs. It encourages system architects to use changing computing paradigms, particularly the widespread use of XML and the rise of RESTful programming into smart city planning. Finally, the OGC views a smart city including layers namely application, business, data, and sensing.

*D. FIWARE*

The FIWARE, also called Open and Agile Smart Cities (OASC), is a generic and open source platform, which has been launched by the European Commission [8]. It aims to develop the core future technologies to make interoperable city services, to provide access to real-time context information, and to implement smart city applications. The adoption of a driven-by-implementation approach is the cornerstone of the FIRWARE. The platform enables developers and communities to create their services based on commonly-defined APIs and data models. The FIWARE defines the following layers: user interfaces, processes, events, innovation, applications, services, interconnectivity, data, intelligence, infrastructure, and stakeholders.

*E. IBM*

The IBM architecture delivers a unified view and underlying technologies for the successful management of cities [9]. The foundational concepts in the IBM architecture are named instrumented, interconnected, and intelligent. These are defined as: *instrumented* refers to sources of real-world data from physical and virtual smart objects, *interconnected* defines the data integrity across computing platforms and the data exchange among various city services. *Intelligent* in this architecture means the inclusion of complex data analytics, modelling, optimizing, and visualizing business processes to make better operational decisions. The IBM enables developers to use city services for the behavior of inhabitants. This allows an effective use of the available physical infrastructure and resources, for instance, in sensing and controlling energy consumption and waste management. The IBM architecture defines the following layers: user interfaces, processes, events, applications, services, interconnectivity, data, intelligence, infrastructure, and stakeholders.

*F. ISO/IEC 30182*

This architecture describes a modelling foundation and ontology alignment for providing interoperability between the system components in different sectors [10]. The architecture defines notions such as organisation, place, metric, service, resource and relationship among these concepts. The ISO/IEC 30182 defines four key types of operational, critical, analytical, and strategic insights regarding data sharing in cities. However, the architecture does not define specific layers, except for a service layer and a partial reference to an interoperability layer.

*G. SOA (Service-oriented architecture model)*

The SOA reference model proposed by Clement provides a holistic architecture for integrating systems such as autonomous vehicles, smart grids, and intelligent traffic management into a smart city [11]. The SOA architecture encapsulates and hides internal mechanisms of city services to enable service interoperability across smart city layers. The architecture defines the three layers: interconnectivity, stakeholders, and city.

*H. Cisco*

The Cisco reference architecture model aims at standardizing concepts and terminologies associated with IoT [12]. The model defines layers including collaboration and processes, application, data abstraction, data accumulation, edge (fog) computing, connectivity, physical devices, and controllers. From the level one (i.e. physical devices and controllers) to the level seven (i.e. collaboration and processes), this model defines functionalities required that should be addressed for a better achievement of IoT initiative values. The Cisco model defines requirements and potentials for enabling IoT smart city architectures.

*I. ESPRESSO (systEmic standardisation apPRoach to Empower Smart citieS and cOmmunities)*

The ESPRESSO proposes a vendor agnostic reference architecture which does not contain any form of technicalities, prescriptive, and certain solutions but it provides cities and communities a mission for implementing enhanced interoperable and standards-based architectures for their specific city context [13]. This architecture aims for improving interoperability maturity and functioning of expanding technology solutions for smart city initiatives. To this end, the ESPRESSO defines key elements and concepts required to be addressed to achieve interoperability between various services within a city and also to increase the interoperability between different cities.

III. CRITERIA-BASED EVALUATION FRAMEWORK

We evaluated the selected architectures presented in the previous section using our proposed evaluation framework presented in Table I. Typically, an evaluation framework is a tool encompassing a collection of criteria that are expected to be satisfied by a product (or service), herein IoT architectures. A product is assessed against the evaluation framework in two steps: (i) the product is scanned for the existence of a feature which is concerned by a criterion, and (ii) an evaluation is carried out signifying the extent to which the product supports that criterion (in our case the result is in the form of scale point) [14]. Our proposed evaluation framework defines a criterion set that is anticipated to be sufficiently addressed by an ideal IoT architecture when dealing with a particular scenario of smart city development. The development of the criteria has been based on an iterative and top/down approach. We identified an initial set of criteria through conducting a systematic literature review of relevant sources such as existing architectures, surveys, reports, and key challenges in implementing IoT-based smart cities. The initial set of criteria was then refined using a set of meta-criteria, i.e. the criteria to assess other criteria, such as being representative, vendor-agnostic, and centred on the features of a smart city architecture to allow soundness, completeness, distinct, and precise. The proposed framework which helps IoT developers to compare and contrast existing architectures is presented in Table 1. Each framework's criterion is accompanied by relevant questions as a guide for the evaluation. The criteria are subsumed under the following groups: (i) criteria related to the architecture profile which assess the high-level features of an architecture (5 criteria), (ii) the criteria related to the architectural layers which examine the architecture's support of generic smart city layers (13 criteria), (iii) criteria related to functional requirements which attest the capabilities of the architecture in addressing key expected IoT functions (9 criteria), and (iv) criteria related to non-functional requirements which check the adherence of the architecture to quality attributes (7 criteria).

TABLE I. THE PROPOSED CRITERIA-BASED EVALUATING FRAMEWORK FOR ANALYSING IoT SMART CITY ARCHITECTURE

| |
|---|
| **Criteria related to architecture profile** |
| **1. Documentation:** Does the architecture have available resources e.g. books, papers, weblog in support of its understanding? |
| **2. Universality:** What geographical regions the proposed architecture can be used to enable a smart city? |
| **3. Application domain:** For what domain of interests (industry sectors) the architecture has been designed? |
| **4. Required expertise:** What background or skills are required to learn the framework? |
| **5. Enabling technologies:** What are the main enabling technologies e.g. cyber-physical systems, big data, and cloud computing used to build the smart city? |
| **Criteria related to architecture layers** |
| **6. User interfaces:** Does the architecture define a layer for showing information to end users? Examples of user interfaces are dashboards, reports, message boards, 3D spaces, 2D maps, and inputs/outputs. |
| **7. Processes:** Does the architecture define a layer for showing business processes/activities in city performing by systems and stakeholders? |
| **8. Events:** Does the architecture define a layer for showing important events in the smart city? Examples of events are such as peak-time vehicle speed, geographic events, local events, and system events. |
| **9. Innovation:** Does the architecture define a layer for representing new business models for city growth and life quality? |
| **10. Applications:** Does the architecture define a layer for showing applications working in the smart city? Examples of application are existing legacy software applications, external software applications, mobile apps, and virtualization. |
| **11. Services:** Does the architecture define a layer for different types of services offered by smart city? Examples of services are e-government services, portals, social services, urban services, etc. |
| **12. Interconnectivity:** Does the architecture define a layer to show addressing interoperability of objects in smart city or IoT architecture? Example of mechanisms/protocols to handle interoperability is using technologies or techniques such as enterprise service bus (ESB), process/workflow orchestrations, adaptors, and wrappers? |
| **13. Data:** Does the architecture define a layer for the various type of data floating across the architecture layers or smart city? Examples of data are real time data from sensors, geospatial data, historical data, and social media. |
| **14. Intelligence:** Does the architecture define a layer for providing sophisticated data analysis? Examples of advanced analysis are adopting APIs for statistical analysis, geospatial analysis, and data analytics? |
| **15. Infrastructure:** Does the architecture define a layer for showing underlying infrastructure of the smart city? Examples of support for having an infrastructure layer is using local connections, Lan, Wan, Man, remote controllers, sensors, actuators, cameras, webcams, smart tags, and RFID. |
| **16. Stakeholders:** Does the architecture define a layer for representing the service provider of smart city and IoT? Examples of service providers in a smart city architecture are project managers, engineers, telecommunication companies, software developers, system architectures, and data architectures. |
| **17. Socio-technical:** Does the architecture define a layer for showing the mechanisms for motivating/stimulating citizen to be a part smart city solution? |
| **18. City:** Does the architecture define a layer for showing the physical city constituents? Examples of city constituents are homes, roads, apartments, parking, public transport, electricity, water cycle, and governance. |
| **Criteria related to functional requirements** |
| **19. Resource discovery:** Does the architecture define the mechanisms or technique (e.g. service discovery protocols) to dynamically and automatically identify new resources (e.g. devices) on the network at any time? |
| **20. Object configuration:** Does the architecture define mechanisms and techniques for setting and connecting objects (e.g. tunnels, gateways, virtual objects) together to communicate in a smart city environment? Does the architecture define reachability policies to resources/network, ports, firewalls and data storages? |
| **21. Interoperability design:** Does the architecture identify potential incompatibilities and corresponding solutions to unify interoperability points to collect, process, and generate data from/to diversity of data sources, legacy devices, and objects? What kind incompatibilities are concerned by the architecture? |
| **22. Data security:** Does the architecture define mechanisms and techniques to keep the security of data and objects across the smart city layers? |
| **23. Data accumulation:** Does the architecture define mechanisms and techniques for the continuous data collection from various objects for processing across all layers? |
| **24. Data cleaning:** Does the architecture define mechanisms/techniques for identifying and correcting inaccurate or incomplete data before storing them into data storages? |
| **25. Data analysis:** Does the architecture implement mechanisms or techniques for identifying useful knowledge and predicting the behaviour of the smart city? |
| **26. Data visualization:** Does the architecture implement mechanisms or techniques for visualizing city data to show its uses? |
| **27. Energy management:** Does the architecture implement mechanisms or techniques in objects such as sensors, actuators, and servers to address the efficient |
| use of resources? |
| **Criteria related to non-functional requirements** |
| **28. Availability:** Does the architecture provides mechanisms and techniques a support for continuous guarantee of obtaining, storing, processing, and providing data and services to users independently of the state of underlying infrastructure? Does the architecture provide clustering mechanisms at the application and service layers? Does the architecture provide redundant storage arrays e.g. RAID (redundant array of independent disks) for the data layer? Does the architecture provide redundant physical links for interconnectivity layer? Does the architecture provide redundant servers for infrastructure layer? |
| **29. Security:** Does the architecture provide mechanisms and techniques for a safe exchanging of data and interactions among technical objects at the application, service, interconnection, data, and infrastructure layers? Does the architecture define an access control list (ACL) on routers, packet filters, firewalls, and network-based intrusion detection systems (IDS) at the network layer? Does the architecture define authentication and authorisation mechanisms at the data user interface, application, service, and data layers? Does the architecture define encryption/decryption mechanism for the data layer? Does the architecture define functional decomposition (separating functional components) to avoid propagating security issues to other architecture components? |
| **30. Interoperability:** Does the architecture provide mechanisms, protocols, and techniques for integration and synergy among heterogeneous objects such as sensor networks, IP network, city data, and human across different platforms to support of interaction/adding/deployment of heterogeneous objects across the layers? |
| **31. Configurability:** Does the architecture define mechanisms and techniques that facilitate architecture configuration to suit new changes? |
| **32. Performance:** Does the architecture define mechanisms and techniques to guarantee best possible throughput of services offering to users with minimum cost? |
| **33. Scalability:** Does the architecture provides mechanisms and techniques to manage increasing the amount of new objects/resources e.g. devices, services and functions to itself to keep expected performance without negatively affecting the quality of existing services in a peak time? Does the architecture define redundant servers to avoid single point of failure? |
| **34. Recoverability:** Does the architecture provide mechanisms and techniques to restore, replace, or fix objects (e.g. services, sensors) that may stop working due to unexpected faults and get the architecture to a state in which it can perform expected functionality? |

## IV. COMPARATIVE ANALYSIS

Table 2 abridges our evaluation results of the selected architectures according to the proposed framework. The evaluation process was performed by reading the available documents of each architecture, its defined layers, guidelines, techniques, and then checking whether the architecture addresses the criteria. For each criterion, we investigated if the architecture defines a mechanism, technique, heuristics, technology, or tool in favor of that criterion. The following delineates the results of our analysis and suggests areas indicating future research directions to improve extant architectures.

**Architecture profile evaluation results.** The results indicate that, except for the ISO/IEC 30182, all the architectures provide publicly available documentation for developers. Unlike the TBS and the ESPRESSO which are offered for enabling smart city initiatives in the UK and Europe with a focus on cultural implications, other architectures seem to be universally applicable at an international level in different cities. The FIWARE is, perhaps, the most suitable for geospatial informatics domain. The existing architectures leverage various technologies to operationalize their functions such as cloud computing, data analytics, crowdsourcing, global position systems (GPS) to name a few. None of the selected architectures specifies the level of sophistication e.g. skills and resources that are required to fully exploit the proposed architecture capabilities in a project by developers.

**Architecture layer evaluation results.** The architectures can be compared in terms of their views to different layers of a typical IoT-based smart city entity. Derived from the existing literature, our framework defines thirteen possible layers which can be considered for a suitable architecture. An overall look at Table 2 shows that the layers namely event, socio-technical, and city have not been sufficiently supported by the selected architectures. More precisely, the event layer, i.e. considering city events such as peak-time vehicle speed, geographic events, local events, and system events, are only supported by the FIRWARE and IBM architectures. The socio-technical layer concerns mechanisms in an IoT architecture to motivate citizens to participate in a smart city development initiative, is addressed by the BSI. Other architectures overlook this criterion. Among the existing proposals, we found that the BSI and ESPRESSO take into account representing physical city constituents e.g. homes, roads, apartments, parking, public transport, electricity, water cycle, and governance as an individual layer in a connection to smart city services.

**Functional requirements evaluation results.** Architectures are expected to address some functional requirements that are critical to the realization of a real smart city. Principally, these requirements are technical centric and are supposed to be implemented in an architecture via enabling technologies, tools, and techniques. In terms of the functional requirements, the framework defines nine criteria. The data management criteria include the data security, data accumulation, data cleaning, data analysis, and data visualization. The criteria related to the data management are sufficiently addressed by the FRAME, IBM, and ESPRESSO where they implement mechanisms and technologies for the data management. On the other hand, the BSI and TSB are weak in providing the data management support.

Dynamically and automatically identifying new resources e.g. devices in the city at any time which is realized using the mechanisms or technique e.g. service discovery protocols is captured by the criterion resource discovery. From the evaluated architectures, only FIWARE address this criterion. The OGC and Cisco, however, define some general guidelines, though a full implementation is not available in their documentation. The argument is made that a more support of resource identification is needed in the existing architectures. The criterion energy management concerns with implementation mechanisms in smart objects such as sensors, actuators, and servers to address the efficient use of resources. None of the existing architectures provides a complete support of the energy management, though the OGC, FIWARE, and ESPRESSO limit themselves to give some general guidelines and skip prescribing mechanisms or techniques. At best, the FIWARE addresses the most number of functional requirements such as resource discovery, object configuration, interoperability design, data security, data accumulation, data cleaning, and data visualization. However, there is a moderate support for data analysis and energy management. The next best architectures are the IBM and ESPRESSO that are considered supportive in addressing functional requirements, though the support of resource discovery and object configuration is not addressed.

**Non-functional requirements evaluation results.** Regarding the adherence of the architectures to the non-functional requirements, the evaluation results reveal that FIWARE and IBM support all the well-known non-functional requirements such as availability, security, interoperability, configurability, performance, scalability, and recoverability, due to adopting enabling technologies in their proposed architectures. In this regard, the OGC and SOA are positioned at second place. On the other hand, the BSI and TSB suffer from a cursory support of quality-related criteria. Interestingly, the fact that the majority of existing architectures define mechanisms to unify incompatibility points among smart objects in order to collect, process, and generate data from/to these varieties of smart objects signifies the importance of the interoperability feature in the existing architectures. The security is concerned across the IoT architecture layers. This implies that the architectures should provide mechanisms for a safe data exchange among objects at the application, service, interconnection, data, and infrastructure layers. For example, the architecture may define an access control list on routers, packet filters, firewalls, and network-based intrusion detection systems (IDS) at the network layer. At the user interface layer, an IoT architecture may define authorization and encryption/decryption mechanisms to access applications, services, and data. In addition, it should be able to avoid propagating security issues to other layers and system components. In the evaluated architectures, the FIRWARE, IBM, and Cisco address the security criterion.

There are a few issues around the comparative analysis using our proposed framework presented in this section. Our review and analysis have been based primarily on the available documents of the selected architecture, and performed at a single point in time. A potential issue with respect to the reliability of the evaluation results (presented in Table II) is the lack of sufficient contextual information in which the architectures have been accommodated. To address this issue, we attempted to find any additional papers and blogs giving more information about the architecture to allow a more accurate assessment. Another issue that may cause threat to the reliability of the presented results is that the evaluation has been made by the authors of this paper. As such, evaluation outcomes may have been exposed to subjectivity or misinterpretation. However, we do not claim generalizability of our analytical results. Further evaluation using more domain experts, for example architecture designers, is needed to reduce the difference in rating of the architectures in favor of the relevant criteria.

## V. RELATED WORK

To the best of our knowledge, there is no explicit criteria-based evaluation framework, as an artefact, in the literature that can be used to characterize the strengths and weaknesses of a typical IoT smart city architecture. We believe the closest studies to this research are the existing literature surveys in which some existing architectures are synthesized to derive a super IoT architecture model. For instance, Santana et al. analysed 23 IoT platforms with respect to enabling technologies as well as functional and non-functional requirements [15]. They derived a reference architecture out of the reviewed architectures. Motivated by the absence of a standardization in the design of smart city architectures, Bastidas et al. identify a list of key requirements related to an IoT architecture [16]. Similarly, the work presented by Kyriazopoulou analyses a few perspectives of the architecture implementation namely layer-based, service-oriented, event-driven, IoT, and then combines the architectures from which a set of basic architectural functional and non-functional requirements are suggested [17]. None of the above mentioned studies can be applicable as a tool to evaluate IoT architectures. A key advantage of our work over other studies is that the proposed framework has measurable criteria in the sense that each criterion has a clear definition, evaluation questions, and scaled format. These measurable criteria enable an assessor in better identification of strengths and shortcomings of the architectures and thus generating a more reliable and traceable evaluation outcome. On the contrary, none of the existing works discuss the way that they have used to assess architectures and left the evaluation rather broad and less articulated. The second advantage of the proposed framework is the research methodology applied to derive the framework. The criteria development has been inspired from the method proposed by Nickerson [18] in which the criteria have been developed and iteratively refined in a top-down/bottom-up fashion. The rigorousness of the criteria development has not been a feature of the earlier studies.

## VI. CONCLUSION

This paper presented a comparison between nine well-known IoT smart city architectures using our proposed evaluation framework. The comparison takes into account a wide range of criteria including layers, functional and non-functional aspects. The evaluation framework and the assessment results may help smart city architecture designer to choose the most suitable standard to use with respect to a particular smart city development initiative. It is important to realize that this paper merely examined the architectures independent from each other. Hence, IoT developers may use a collection of elements in different existing architectures in a hybrid manner. Our further work is to extend the criterion set and compare many other existing IoT architectures in order to obtain an overall assessment of the current body of knowledge in the field.

TABLE II. COMPARATIVE ANALYSIS RESULTS

NS: Not stated, ●: Completely addressed across all layers, ◐: Considerably addressed in some layers, ◑: Moderately addressed (some guidelines for the support of interoperability but details are lacking), ◔: Slightly addressed (merely some referrals), ○: Not at all

| Criterion | BSI | TBS | QGC | FIWARE | IBM | ISO/IEC 30182 | SOA | Cisco | ESPRESSO |
|---|---|---|---|---|---|---|---|---|---|
| **Architecture profile related criteria** | | | | | | | | | |
| Documentation | Yes | Yes | Yes | Yes | Yes | No | Yes | Yes | Yes |
| Universality | International | Country (UK) | NS | International | International | International | International | International | Europe |
| Application domain | NS | In general | Geospatial informatics | In general | In general | In general | In general | In general | In general |
| Enabling technologies | SOA | NS | Cloud computing, crowdsourcing, data analytics | cloud computing, data analytics, OpenStack | Cloud computing, data analytics | NS | SOA, cloud computing | Cloud computing, edge computing, data analytics | TOGAF, Crowdsourcing |
| Required expertise | NS | NS | NS | NS | NS | NS | NS | NS | NS |
| **Architecture layer related criteria** | | | | | | | | | |
| User interfaces | ● | ◔ | ◑ | ● | ● | ○ | ◔ | ○ | ○ |
| Processes | ○ | ○ | ● | ● | ● | ○ | ◑ | ● | ● |
| Events | ○ | ○ | ○ | ● | ● | ◑ | ◔ | ○ | ◔ |
| Innovation | ◔ | ◑ | ○ | ○ | ○ | ○ | ○ | ○ | ○ |
| Applications | ○ | ◐ | ● | ● | ● | ○ | ◑ | ● | ● |
| Services | ● | ● | ● | ● | ● | ● | ◑ | ● | ● |
| Interconnectivity | ● | ◐ | ● | ● | ● | ◑ | ◐ | ● | ● |
| Data | ● | ◐ | ● | ● | ● | ○ | ◑ | ● | ● |
| Intelligence | ● | ◐ | ● | ● | ● | ○ | ◔ | ● | ● |
| Infrastructure | ● | ● | ● | ● | ● | ○ | ◑ | ● | ● |
| Stakeholders | ● | ● | ○ | ● | ● | ◑ | ◐ | ◔ | ● |
| Socio-technical | ● | ◔ | ○ | ○ | ○ | ○ | ◔ | ◔ | ● |
| City | ● | ○ | ◔ | ○ | ○ | ○ | ◐ | ○ | ● |
| **Functional requirement related criteria** | | | | | | | | | |
| Resource discovery | ○ | ○ | ◑ | ● | ○ | ○ | ○ | ◑ | ○ |
| Object configuration | ○ | ○ | ○ | ● | ○ | ○ | ○ | ◑ | ○ |
| Data security | ◔ | ◑ | ◔ | ● | ● | ◑ | ◔ | ◑ | ◔ |
| Data accumulation | ◔ | ○ | ● | ● | ● | ○ | ◔ | ● | ● |
| Data cleaning | ○ | ○ | ◔ | ● | ● | ○ | ◔ | ● | ● |
| Data analysis | ◔ | ○ | ● | ● | ● | ○ | ◔ | ● | ● |
| Data visualization | ◔ | ○ | ● | ● | ● | ○ | ◔ | ◔ | ● |
| Energy management | ◔ | ◑ | ◑ | ◑ | ◔ | ○ | ◔ | ○ | ◑ |
| **Non-functional requirement related criteria** | | | | | | | | | |
| Availability | ○ | ◔ | ● | ● | ● | ◔ | ◔ | ● | ○ |
| Security | ◔ | ◔ | ◔ | ● | ● | ◑ | ◔ | ● | ● |
| Interoperability | ◔ | ◔ | ● | ● | ● | ● | ● | ◑ | ● |
| Configurability | ○ | ○ | ○ | ● | ● | ○ | ◑ | ○ | ○ |
| Performance | ○ | ◔ | ○ | ● | ● | ○ | ◔ | ◔ | ○ |
| Scalability | ○ | ◔ | ◔ | ● | ● | ○ | ◔ | ◔ | ○ |
| Recoverability | ○ | ○ | ◔ | ● | ● | ○ | ◔ | ○ | ○ |

ACKNOWLEDGMENT

We would like to thank and acknowledge the support of Australian Federal Government Smart Cities and Suburbs program, IoT Alliance Australia and the Knowledge Economy Institute at UTS.